\documentclass{article}

\usepackage{arxiv}

\usepackage[utf8]{inputenc} 
\usepackage[T1]{fontenc}    
\usepackage{hyperref}       
\usepackage{url}            
\usepackage{booktabs}       
\usepackage{amsfonts}       
\usepackage{amssymb}        
\usepackage{amsmath}        
\usepackage{nicefrac}       
\usepackage{microtype}      
\usepackage{lipsum}		
\usepackage{graphicx}
\usepackage[numbers]{natbib}
\usepackage{doi}
\usepackage{wrapfig}

\title{Sparse arrays in R: the spray package}

\author{ \href{https://orcid.org/0000-0001-5982-0415}{\includegraphics[width=0.03\textwidth]{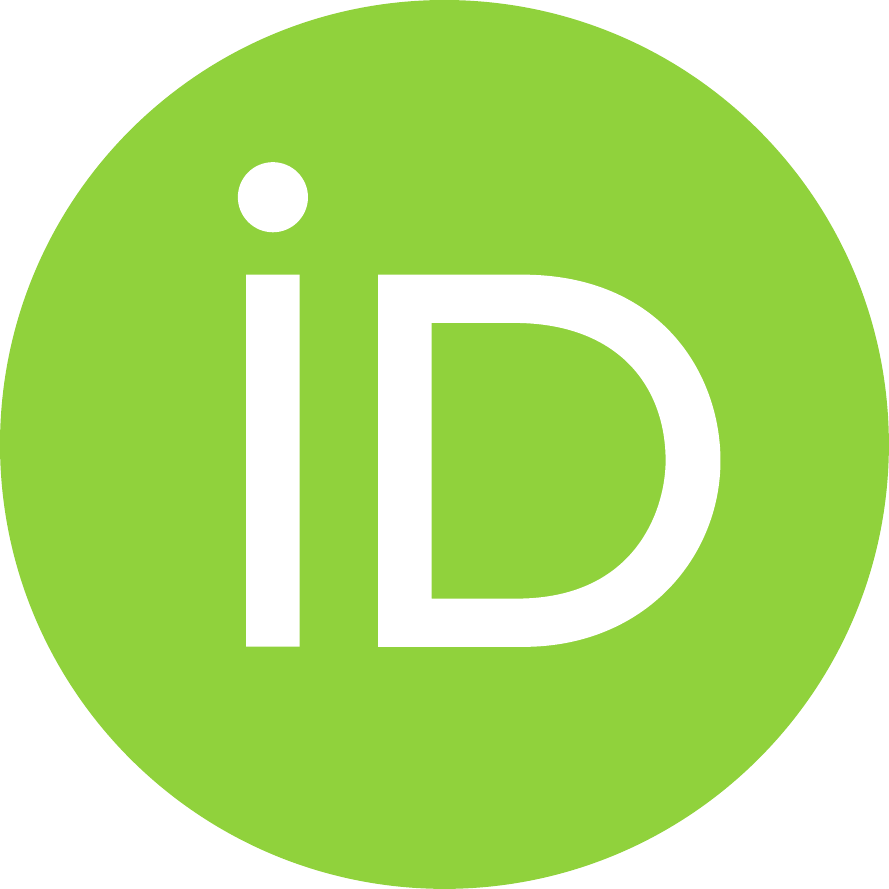}\hspace{1mm}Robin K. S.~Hankin}\thanks{\href{https://academics.aut.ac.nz/robin.hankin}{work};  
\href{https://www.youtube.com/watch?v=JzCX3FqDIOc&list=PL9_n3Tqzq9iWtgD8POJFdnVUCZ_zw6OiB&ab_channel=TrinTragulaGeneralRelativity}{play}} \\
 Auckland University of Technology\\
	\texttt{hankin.robin@gmail.com} \\
}

\renewcommand{\shorttitle}{The \textit{spray} package}

\hypersetup{
pdftitle={Sparse arrays in R},
pdfsubject={q-bio.NC, q-bio.QM},
pdfauthor={Robin K. S.~Hankin},
pdfkeywords={Sparse arrays}
}

\usepackage{Sweave}
\begin{document}
\maketitle

\begin{abstract}
  In this short article I introduce the {\tt spray} package, which
  provides some functionality for handling sparse arrays.  The package
  uses the~{\tt C++} Standard Template Library's {\tt map}
  class to store and retrieve elements.  One
  natural application for sparse arrays is multivariate polynomials
  and I give two examples of the package in use, one drawn from the
  fields of random walks on lattices and one from the field of
  recreational combinatorics.  The package is available on CRAN at
  \url{https://CRAN.R-project.org/package=spray}.
\end{abstract}
\keywords{Multivariate polynomials, sparse arrays}

\section{Introduction}

\setlength{\intextsep}{0pt}
\begin{wrapfigure}{r}{0.2\textwidth}
  \begin{center}
\includegraphics[width=1in]{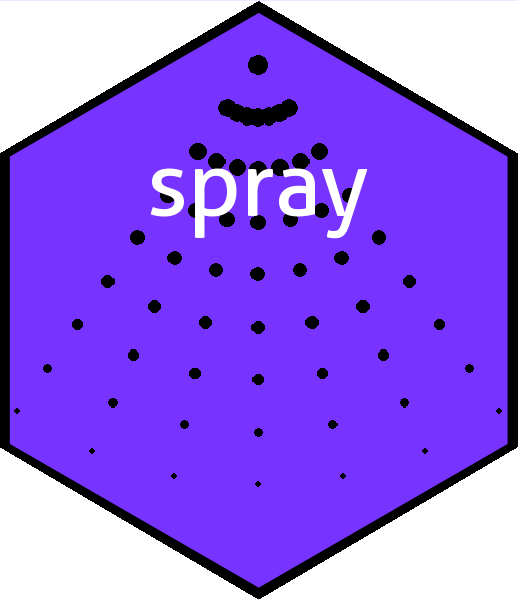}
  \end{center}
\end{wrapfigure}

The {\tt multipol} package~\citep{hankin2008} furnishes the
{\tt R} programming language with functionality for multivariate
polynomials.  However, the {\tt multipol} package was noted as being
inefficient in many common cases: the package stores multivariate
polynomials as arrays and this often involves storing many zero
elements which consume computational and memory resources
unnecessarily.

One suggestion was to use sparse arrays---in which nonzero elements
are stored along with an index vector describing their
coordinates---instead of arrays.  In this short document I introduce
the {\tt spray} package which provides functionality for sparse arrays
and interprets them as multivariate polynomials.  Some of the
underlying design philosophy is discussed in the appendix.  Here,
`sparse multinomial' is defined as one whose array representation is
sufficiently sparse to make taking advantage of is sparseness
worthwhile.  However, other definitions of sparseness may be useful
and I outline some below.

\subsection{Existing work}

The~{\tt slam} package~\citep{hornik2014} provides some sparse array
functionality but is not intended to interpret arbitrary dimensional
sparse arrays as multivariate polynomials; the {\tt rSympy} package
does not, as of 2017, implement sparse multivariate polynomials.  The
{\tt mpoly}~\citep{kahle2013} package handles multivariate polynomials
but does not accept negative powers, nor is it designed for
efficiently processing large multivariate polynomials; I present some
timings below.  The {\tt mpoly} package is different in philosophy
from both the~{\tt spray} package and~{\tt multipol} in
that~{\tt mpoly} is more ``symbolic'' in the sense that it
admits---and handles appropriately---named variables, whereas my
packages do not make any reference to the {\em names} of the
variables.  As Kahle points out, naming the variables
allows a richer and more natural suite of functionality;
straightforward {\tt mpoly} idiom is somewhat strained in {\tt spray}.

(The {\tt mvp} package~\citep{hankin2019} is now available; this uses
a more powerful concept of sparsity).

\section{Sparse arrays}

Base R has extensive support for multidimensional arrays.  Consider

\begin{Schunk}
\begin{Sinput}
R> a <- array(1, dim=2:5)
\end{Sinput}
\end{Schunk}

\noindent The resulting object requires storage of~$2\times 3\times
4\times 5=120$ floating point numbers, which are represented in an
elegant format amenable to Cartesian extraction and replacement.
However, arrays in which many of the elements are zero are common and
in this case storing only the nonzero elements and their positions
would be a more compact and efficient representation.  To create a
sparse array object in the {\tt spray} package, one specifies a matrix
of indices {\tt M} with each row corresponding to the position of a
nonzero element, and a numeric vector of values:

\begin{Schunk}
\begin{Sinput}
R> library("spray")
R> M <- matrix(c(0, 0, 0, 1, 0, 0, 1, 1, 1, 2, 0, 3), ncol=3)
R> M
\end{Sinput}
\begin{Soutput}
     [,1] [,2] [,3]
[1,]    0    0    1
[2,]    0    0    2
[3,]    0    1    0
[4,]    1    1    3
\end{Soutput}
\begin{Sinput}
R> S1 <- spray(M, 1:4)
R> S1
\end{Sinput}
\begin{Soutput}
           val
 0 0 1  =    1
 0 1 0  =    3
 0 0 2  =    2
 1 1 3  =    4
\end{Soutput}
\end{Schunk}

Thus {\tt S1[0,0,2] = 2}.  The representation of the spray object
does not preserve the order of the index rows in the argument,
although a particular index row is associated unambiguously with a
unique numeric value.  This is because the {\tt STL} map class
does not preserve the orders of its elements.  This does not matter,
as the order in which the elements are stored is immaterial in the
use-cases presented here.  

Extract and replace methods require the index to be a
matrix\footnote{Indexing with a vector (interpreted as a row of the
index matrix) is problematic.  The package requires idiom such
as~{\tt S[1,2,3]} and~{\tt S[1,1:3,3]} to work as expected; and
because {\tt [.spray()} and {\tt [<-.spray()} dispatch on the first
argument, the package does not attempt to guess what the user
intended.}:
\begin{Schunk}
\begin{Sinput}
R> S1[diag(3)] <- -3
R> S1
\end{Sinput}
\begin{Soutput}
           val
 0 0 1  =   -3
 0 1 0  =   -3
 0 0 2  =    2
 1 1 3  =    4
 1 0 0  =   -3
\end{Soutput}
\end{Schunk}

\noindent We can see that a value with an existing index is
overwritten, while new elements are created as necessary.  Addition is
implemented:

\begin{Schunk}
\begin{Sinput}
R> M2 <- matrix(c(
+      6, -7,  8,
+      0,  0,  2,
+      1,  1,  3), byrow=TRUE, ncol=3)
R> S2 <- spray(M2, c(17, 11 ,-4))
R> S2
\end{Sinput}
\begin{Soutput}
            val
 6 -7 8  =   17
 0  0 2  =   11
 1  1 3  =   -4
\end{Soutput}
\begin{Sinput}
R> S1 <- S1 + S2
R> S1
\end{Sinput}
\begin{Soutput}
            val
 0  0 1  =   -3
 0  1 0  =   -3
 0  0 2  =   13
 1  0 0  =   -3
 6 -7 8  =   17
\end{Soutput}
\end{Schunk}

\noindent Thus element~{\tt [0,0,2]} becomes $2+11=13$, while
element~{\tt [1,1,3]} cancels and thus vanishes.  There is no
requirement for indices to be positive: Element~{\tt [6,-7,8]} is new
(with value~17).  Even though the representation of the spray object
does not preserve the order of the index rows in the argument, a
particular index row is associated unambiguously with a unique numeric
value.

\section{The spray package and multivariate polynomials}

One natural application for {\tt spray} objects is multivariate
polynomials~\citep{hankin2008}.  I will first discuss the
univariate case and then progress to multivariate polynomials.

\subsection{Univariate polynomials}
Univariate polynomials are a good place to start.  Suppose the
polynomial

\[
A = 1+2x^3 + 6x^8
\]

\noindent were to be represented using R objects.  One natural
approach, taken in the {\tt polynomial} package~\citep{venables2016},
is to store the coefficients in a vector:

\begin{Schunk}
\begin{Sinput}
R> library("polynom")
R> A <- polynomial(c(1, 0, 0, 2, 0, 0, 0, 0, 6))
R> dput(A)
\end{Sinput}
\begin{Soutput}
structure(c(1, 0, 0, 2, 0, 0, 0, 0, 6), class = "polynomial")
\end{Soutput}
\begin{Sinput}
R> A
\end{Sinput}
\begin{Soutput}
1 + 2*x^3 + 6*x^8 
\end{Soutput}
\end{Schunk}

\noindent But again see how the {\tt R} object thus created
stores zero elements, which can be problematic if the polynomial in
question is large degree and sparse.  Similar issues arise in the case
of multivariate polynomials.  The~{\tt multipol}
package~\citep{hankin2008} uses a similar methodology, storing
coefficients as an arbitrary-dimensional array.  However, as noted
above, this often leads to inefficient computation.

\subsection{Multivariate polynomials}

One natural and useful interpretation of a sparse array is as a
multivariate polynomial.  Consider the following sparse array:

\begin{Schunk}
\begin{Sinput}
R> S3 <- spray(matrix(c(0,0,0, 0,0,1, 1,1,1, 3,0,0), byrow=TRUE, ncol=3), 1:4)
R> S3
\end{Sinput}
\begin{Soutput}
           val
 0 0 0  =    1
 0 0 1  =    2
 1 1 1  =    3
 3 0 0  =    4
\end{Soutput}
\end{Schunk}

\noindent It is natural to interpret the rows of the index matrix as
powers of different variables of a multivariate polynomial, and the
values as being the coefficients.

This is realized in the package using the {\tt polyform} option,
which if set to {\tt TRUE}, modifies the print method:

\begin{Schunk}
\begin{Sinput}
R> options(polyform = TRUE)
R> S1
\end{Sinput}
\begin{Soutput}
-3*z -3*y +13*z^2 -3*x +17*x^6*y^-7*z^8
\end{Soutput}
\end{Schunk}

\noindent (only the print method has changed; {\tt S1} is as before).
The print method interprets, by default, the three columns as
variables $x,y,z$ although this behaviour is user-definable.  With
this interpretation, multiplication and addition have natural
definitions as multivariate polynomial multiplication and addition:

\begin{Schunk}
\begin{Sinput}
R> S1 + S2
\end{Sinput}
\begin{Soutput}
-3*z -3*y +24*z^2 -4*x*y*z^3 -3*x +34*x^6*y^-7*z^8
\end{Soutput}
\begin{Sinput}
R> S1 * S2
\end{Sinput}
\begin{Soutput}
+12*x^2*y*z^3 -51*x^7*y^-7*z^8 +289*x^12*y^-14*z^16
-68*x^7*y^-6*z^11 -33*y*z^2 -52*x*y*z^5 +143*z^4 +12*x*y*z^4
+12*x*y^2*z^3 +408*x^6*y^-7*z^10 -51*x^6*y^-6*z^8 -33*x*z^2
-33*z^3 -51*x^6*y^-7*z^9
\end{Soutput}
\begin{Sinput}
R> S1^2
\end{Sinput}
\begin{Soutput}
+442*x^6*y^-7*z^10 +9*y^2 +18*y*z +9*z^2 +18*x*y -78*z^3
+18*x*z +169*z^4 +9*x^2 -102*x^6*y^-7*z^9 -78*y*z^2
-102*x^6*y^-6*z^8 -78*x*z^2 -102*x^7*y^-7*z^8
+289*x^12*y^-14*z^16
\end{Soutput}
\end{Schunk}

It is possible to introduce an element of symbolic calculation,
exhibiting familiar algebraic identities.  Consider the {\tt lone()}
function, which creates a sparse array whose multivariate polynomial
interpretation is a single variable:

\begin{Schunk}
\begin{Sinput}
R> x <- lone(1, 3)
R> y <- lone(2, 3)
R> z <- lone(3, 3)
R> options(polyform = FALSE)
R> list(x, y, z)
\end{Sinput}
\begin{Soutput}
[[1]]
           val
 1 0 0  =    1

[[2]]
           val
 0 1 0  =    1

[[3]]
           val
 0 0 1  =    1
\end{Soutput}
\begin{Sinput}
R> options(polyform = TRUE)
R> (1 + x + y)^3
\end{Sinput}
\begin{Soutput}
1 +3*x^2 +3*y +x^3 +3*x +3*x^2*y +6*x*y +3*x*y^2 +3*y^2 +y^3
\end{Soutput}
\begin{Sinput}
R> (x + y) * (y + z) * (x + z) - (x + y + z) * (x*y + x*z + y*z)
\end{Sinput}
\begin{Soutput}
-x*y*z
\end{Soutput}
\begin{Sinput}
R> (x + y) * (x - y) - (x^2 - y^2)
\end{Sinput}
\begin{Soutput}
the NULL multinomial of arity 3
\end{Soutput}
\end{Schunk}

\subsection{The null polynomial and arity issues}

The package is intended to provide functionality for sparse arrays,
one interpretation of which is multivariate polynomials.  The package,
implementing sparse arrays, forbids the addition of two sparse arrays
with different dimensionalities:

\begin{Schunk}
\begin{Sinput}
R>  lone(1, 2) + lone(1, 1)
\end{Sinput}
\begin{Soutput}
Error: arity(S1) == arity(S2) is not TRUE
\end{Soutput}
\end{Schunk}

One problematic object is the empty array.  Zero multinomials are
represented with a zero-row index matrix and zero-length numeric
vector of values.  Because a spray object is a sparse array, a zero
multinomial must have a specific arity\footnote{This philosophy is
  different from earlier versions of the software which treated the
  empty array as the zero multinomial, with which addition and
  multiplication were defined algebraically.  I would like to thank an
  anonymous \emph{R Journal} referee for this insight.}.

\subsection{Algebraic identities}

Similar but more involved techniques can be used to prove Euler's
four-square identity and Degen's eight-square identity, given in the
package's test suite.  However, it should be noted that the
{\tt mpoly} package has a more natural idiom and does not suffer from
the visual defect of arbitrary term ordering.

\subsection{Further functionality}

Multivariate polynomials have a natural interpretation as functions:
  
\begin{Schunk}
\begin{Sinput}
R> (S4 <- spray(cbind(1:3, 3:1), 1:3))
\end{Sinput}
\begin{Soutput}
+x*y^3 +2*x^2*y^2 +3*x^3*y
\end{Soutput}
\begin{Sinput}
R> f <- as.function(S4)
R> f(c(1, 2))
\end{Sinput}
\begin{Soutput}
 X 
22 
\end{Soutput}
\end{Schunk}

The last line showing the result of substituting~$x=1,y=2$
into~{\tt S4}.  Other algebraic operations include substitution and
partial differentiation.  Consider the homogeneous polynomial in three
variables:

\begin{Schunk}
\begin{Sinput}
R> (S5 <- homog(3, 3))
\end{Sinput}
\begin{Soutput}
+x^2*y +y^3 +x^2*z +x^3 +x*y^2 +x*y*z +y^2*z +x*z^2 +z^3
+y*z^2
\end{Soutput}
\end{Schunk}

Interpreting~{\tt S5} as a multivariate polynomial with
variables~$x,y,z$ we may substitute~$y=5$ using the {\tt subs()}
function:

\begin{Schunk}
\begin{Sinput}
R> subs(S5, 2, 5)
\end{Sinput}
\begin{Soutput}
125 +5*x^2 +x^3 +y^3 +x^2*y +25*x +5*x*y +25*y +x*y^2 +5*y^2
\end{Soutput}
\end{Schunk}

Differentiation is also straightforward.  Suppose we wish to calculate
the multivariate polynomial corresponding to
 
\[
\frac{\partial^6}{\partial x\,\partial^2y\,\partial^3z}
\left(xyz + x+2y+3z\right)^3
\]

This would be

\begin{Schunk}
\begin{Sinput}
R> aderiv((xyz(3) + linear(1:3))^3, 1:3)
\end{Sinput}
\begin{Soutput}
+216*x +108*x^2*y
\end{Soutput}
\end{Schunk}

\section{Extraction and manipulation of coefficients}

The {\tt spray} package uses {\tt disordR}
discipline~\cite{hankin2022_disordR_arxiv} for access and manipulation
of coefficients.

\begin{Schunk}
\begin{Sinput}
R> options(polyform = FALSE)
R> (X <- rspray())
\end{Sinput}
\begin{Soutput}
           val
 2 1 2  =    8
 0 2 2  =    2
 0 0 1  =    8
 1 2 1  =    4
 2 1 0  =    6
 1 2 2  =    8
 0 0 2  =    9
\end{Soutput}
\end{Schunk}

Above, we note that the terms of {\tt X} are held in an
implementation-specific order, as it uses the {\tt STL } map
class~\cite{musser2009}.  Thus extraction of the coefficients cannot
return an ordinary R vector, as ordinary R vectors are stored in a
uniquely-specified order.  The coefficients of {\tt X} are retrieved
using function {\tt coeffs()}:

\begin{Schunk}
\begin{Sinput}
R> coeffs(X)
\end{Sinput}
\begin{Soutput}
A disord object with hash 79193703471540ec10f921b46d470cd5c0a1740b and elements
[1] 8 2 8 4 6 8 9
(in some order)
\end{Soutput}
\end{Schunk}

Function {\tt coeffs()} returns an object of class {\tt disord}, which
is designed to allow for an implementation-specific order of the terms
of the {\tt spray} object.  The {\tt disordR} and {\tt mvp}
packages~\cite{hankin2019} include an extensive discussion.

\section{The package in use: some examples}

Multivariate polynomials are useful and efficient structures in a
variety of applications.  Here I give two examples of the package in
use: One drawn from the field of random walks on lattices, and one
from recreational combinatorics.

\subsection{Random walks on lattices}

Random walks on periodic lattices find application in a wide range of
applied mathematics including the study of molecular and ionic
crystals~\citep{hollander1982}, polymers~\citep{scheunders1989} and
photosynthetic units~\citep{montroll1969}.  The basic idea is that
some entity (exciton, ion, etc) has a well-defined position on a
periodic lattice~$\left(\mathbb{Z}/n\mathbb{Z}\right)^d$; it then
moves on the lattice, performing a random walk.  The examples here
have~$d=2$ but extension to arbitrary dimensions is immediate.

The periodic lattice itself may be identified with a multivariate
polynomial in~$d$ variables, here~$x$ and~$y$; the probability of the
entity being at point~$(n,m)$ is the coefficient of~$x^ny^m$.

The entity typically moves between adjacent nodes according to a
\emph{kernel} polynomial whose coefficients are the probabilities of
the moves.  Periodicity may be enforced simply by wrapping the
polynomial using modular arithmetic.

In many cases, entities are not necessarily conserved: the entity may
decay (usually with a fixed probability per timestep), or be
annihilated when it encounters a particular node in the lattice (a
`trap'~\citep{scheunders1989}, corresponding to the formation of sugar
in the cell).  Here, we work on a~$17\times 17$ lattice as a large but
computationally tractable domain.

All these processes have natural and efficient~{\tt R} idiom in
the {\tt spray} package.  We may specify a kernel allowing movement to
adjacent nodes, or to stay in the same place with equal probability:

\begin{Schunk}
\begin{Sinput}
R> d <- 2
R> kernel <- spray(rbind(0, diag(d), -diag(d)))/(1 + 2*d)
\end{Sinput}
\end{Schunk}

\noindent At the first timestep, the the entity is at, say,
point~$\left(10,10\right)$ with probability 1:

\begin{Schunk}
\begin{Sinput}
R> initial <- spray(rep(10, d))
\end{Sinput}
\end{Schunk}

\noindent Finding the probability mass function of the entity after,
say 14 timesteps, is straightforward:

\begin{Schunk}
\begin{Sinput}
R> t14 <- initial * kernel^14
\end{Sinput}
\end{Schunk}

Traps may be assigned using standard indexing and we will work with
a~$17\times 17$ array:

\begin{Schunk}
\begin{Sinput}
R> traps <- matrix(c(2, 3, 3, 5), 2, 2)
R> n <- 17
\end{Sinput}
\end{Schunk}

Then we may calculate the evolution of the probability mass function as follows:

\begin{Schunk}
\begin{Sinput}
R> timestep <- function(state, kernel, traps){
+    state <- state * kernel
+    state <- spray(index(state)%%n, coeffs(state), addrepeats = TRUE)
+    state[traps] <- 0
+    return(state)
+  }
\end{Sinput}
\end{Schunk}

In function {\tt timestep()}, the first line uses standard
multivariate polynomial multiplication to advance the state of the
entity; the second enforces periodic boundary conditions, and the
third implements the traps' annihilation of the entity.  The
probability of the entity still existing after 100 timesteps is then:

\begin{Schunk}
\begin{Sinput}
R> state <- initial
R> for(i in 1:100){state <- timestep(state, kernel, traps)}
R> sum(coeffs(state))
\end{Sinput}
\begin{Soutput}
[1] 0.9006642
\end{Soutput}
\end{Schunk}

Note the streamlined {\tt R} idiom: It is not clear how such
manipulations could be performed using the {\tt mpoly} or the
{\tt multipol} packages.

\subsection{Recreational combinatorics}

Suppose we consider a chess knight and ask how many ways are there for
the knight to return to its starting square in 6 moves.  Such
questions are most naturally answered by using generating functions.

On an infinite chessboard, we might define the multivariate generating
polynomial\footnote{Standard terminology, although it might be more
  accurately referred to as a multivariate Laurent polynomial.}~$k$
for a knight as

\[
k = x^{2}y + x^2y^{-1} + x^{-2}y + x^{-2}y^{-1} + xy^{2} + xy^{-2} + x^{-1}y^{2} + x^{-1}y^{-2} 
\]

where we have identified powers of~$x$ with squares moved horizontally
(counted algebraically, negative powers mean move to the left), and
powers of~$y$ with squares moved vertically.  Then the coefficient
of~$x^{a}y^{b}$ in~$k$ is the number of ways of moving from the origin
[that is, $x^{0}y^{0}$] to square~$(a,b)$.  Similarly, $k^n$ is the
generating function for a knight which makes~$n$ moves: The
coefficient of~$x^{a}y^{b}$ in~$k^n$ is the number of ways of moving
from the origin to square~$(a,b)$.

The {\tt R} idiom for this is straightforward; we define
{\tt chess\_knight}, a spray object with rows corresponding to the
possible moves the chess piece may make:
  
\begin{Schunk}
\begin{Sinput}
R> chess_knight <- 
+    spray(matrix(
+        c(1, 2, 1, -2, -1, 2, -1, -2, 2, 1, 2, -1, -2, 1, -2, -1),
+        byrow = TRUE,ncol = 2))
R> options(polyform = FALSE)
R> chess_knight
\end{Sinput}
\begin{Soutput}
           val
  1  2  =    1
  1 -2  =    1
 -1  2  =    1
 -1 -2  =    1
  2  1  =    1
  2 -1  =    1
 -2  1  =    1
 -2 -1  =    1
\end{Soutput}
\begin{Sinput}
R> options(polyform = TRUE)
R> chess_knight
\end{Sinput}
\begin{Soutput}
+x*y^2 +x*y^-2 +x^-1*y^2 +x^-1*y^-2 +x^2*y +x^2*y^-1 +x^-2*y
+x^-2*y^-1
\end{Soutput}
\end{Schunk}

\noindent Then {\tt chess\_knight[i,j]} gives the number of ways the
piece can move from square~{\tt [0,0]} to~{\tt [i,j]};
and~{\tt (chess\_knight\^{}n)[i,j]} gives the number of ways the
piece can reach~{\tt [i,j]} in {\tt n} moves.  To calculate the
number of ways that the piece can return to its starting square we
simply raise {\tt chess\_knight} to the sixth power and extract the
{\tt [0,0]} coefficient:
\begin{Schunk}
\begin{Sinput}
R> constant(chess_knight^6, drop = TRUE)
\end{Sinput}
\begin{Soutput}
[1] 5840
\end{Soutput}
\end{Schunk}

(function {\tt constant()} extracts the coefficient corresponding to
zero power).  One natural generalization would be to arbitrary
dimensions.  A d-dimensional knight moves two squares in one
direction, followed by one square in another direction:

\begin{Schunk}
\begin{Sinput}
R> knight <- function(d){
+    n <- d * (d - 1)
+    out <- matrix(0, n, d)
+    out[cbind(rep(seq_len(n), each=2), c(t(which(diag(d)==0, arr.ind=TRUE))))] <- seq_len(2)
+    spray(rbind(out, -out, `[<-`(out, out==1, -1),`[<-`(out, out==2, -2)))
+  }
\end{Sinput}
\end{Schunk}

Then, considering a four-dimensional chessboard
(Figure~\ref{four_dimensional_knight}):

\begin{figure}[h]
  \centering \includegraphics[width=10cm]{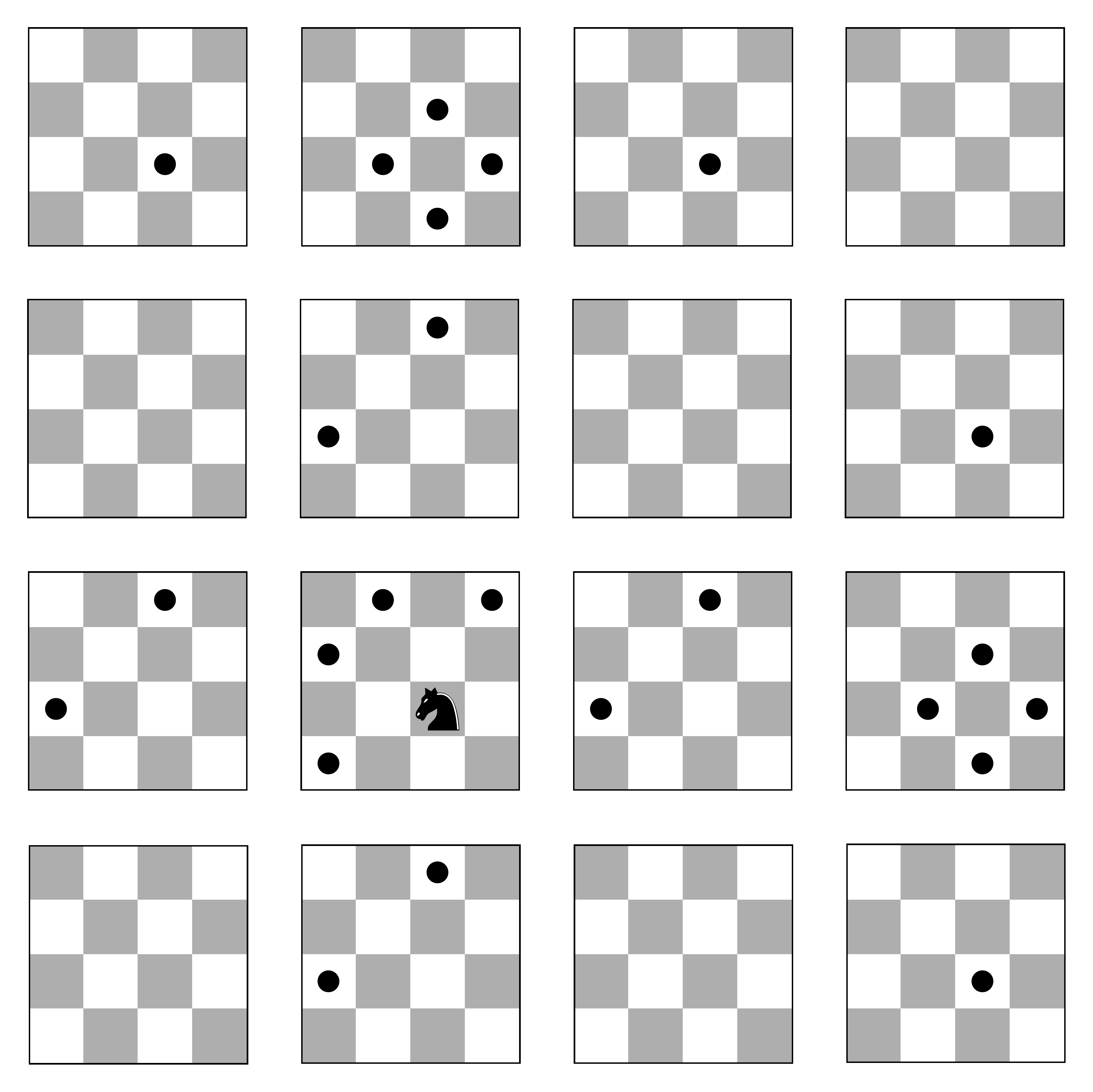}
  \caption{Four-dimensional knight\label{four_dimensional_knight} on a $4\times 4\times 4\times 4$ board.  Cells attacked by the knight shown by dots}
\end{figure}

\begin{Schunk}
\begin{Sinput}
R> constant(knight(4)^6, drop = TRUE)
\end{Sinput}
\begin{Soutput}
[1] 10117920
\end{Soutput}
\end{Schunk}

It is in such cases that the efficiency of the {\tt map} class
becomes evident: On my system (3.4\,GHz Intel Core i5 iMac), the
above call took just under~0.4 seconds of elapsed time whereas the
same\footnote{Because {\tt mpoly} does not accept negative powers, the
  calculation was equivalent to {\tt (knight(4) + xyz(4)\^{}2)\^{}6}.
  Also note that the {\tt multipol} package is not able to execute
  these commands in a reasonable time.} calculation took over 173
seconds using {\tt mpoly}.

If we want the number of ways to return to the starting point in 6 or
fewer moves, we can simply add the unit multinomial and take the sixth
power of the sum:
  
\begin{Schunk}
\begin{Sinput}
R> constant((1 + knight(4))^6, drop=TRUE)
\end{Sinput}
\begin{Soutput}
[1] 10306561
\end{Soutput}
\end{Schunk}
  
(0.6 seconds for~{\tt spray} vs 275 seconds for~{\tt mpoly}).  For 8
moves, the differences are more pronounced, with {\tt spray} taking
4.0 seconds and~{\tt mpoly} requiring more than 1500 seconds).

\section{Conclusions and further work}

The {\tt spray} package provides functionality for sparse,
arbitrarily-dimensioned arrays.  One natural interpretation of a
sparse array is as a multivariate polynomial and the package leverages
the {\tt map} class of {\tt C++} to give fast polynomial
multiplication.

The functionality provided overlaps with that of {\tt multipol} and
{\tt mpoly}.  The {\tt multipol} package is too slow to be of
practical value for any but the smallest illustrative objects.

The different name-based philosophy employed by the {\tt mpoly}
package is certainly an advantage in terms of natural {\tt R}
idiom, although there is a performance penalty.  There are also
occasional applications of multivariate polynomials (such as random
walks on lattices) in which the structure of {\tt spray} is a
conceptual advantage.

British mathematician J. H. Wilkinson famously defined a matrix to be
``sparse'' if it has enough zeros that it pays to take advantage of
them; this definition applies to the arrays considered here.  However,
other definitions of sparsity are possible.  Consider the following
example, taken from \cite{kahle2013}:

\[
ab^2+bc^2+cd^2+\cdots+yz^2+za^2
\].

The {\tt spray} idiom for such an expression is

\begin{Schunk}
\begin{Sinput}
R> a <- diag(26)
R> options(sprayvars = letters)
R> a[1 + cbind(0:25, 1:26) %% 26] <- 2
R> spray(a)
\end{Sinput}
\begin{Soutput}
+r*s^2 +c*d^2 +f*g^2 +e*f^2 +d*e^2 +g*h^2 +i*j^2 +h*i^2 +j*k^2
+y*z^2 +k*l^2 +b*c^2 +s*t^2 +l*m^2 +m*n^2 +w*x^2 +n*o^2 +a*b^2
+v*w^2 +p*q^2 +q*r^2 +t*u^2 +o*p^2 +u*v^2 +x*y^2 +a^2*z
\end{Soutput}
\end{Schunk}

\noindent ---but it is clear that the index matrix has a large degree
of sparseness which {\tt mpoly} takes advantage of and {\tt spray}
does not.  Further work might include the development of name-based
multivariate polynomials using concepts from {\tt STL} to provide
the best of both worlds (and indeed the {\tt mvp}
package~\citep{hankin2019} does just this, and more).

\bibliographystyle{plain}
\bibliography{spray_arxiv}

\newpage
\appendix
\section{Package philosophy}

The {\tt spray} package does not interact with or depend on {\tt
  multipol} in any way, owing to the very different design
philosophies used.  The package uses the~{\tt C++} Standard Template
Library's {\tt map} class~\citep{musser2009} to store and retrieve
elements.

A \emph{map} is an associative container that stores values indexed by
a key, which is used to sort and uniquely identify the values.  In the
package, the key is a {\tt vector} object or a {\tt deque}
object with (signed) integer elements.

In the {\tt STL}, a {\tt map} object stores keys and
associated values in whatever order the software considers to be most
propitious.  This allows faster access and modification times but the
order in which the maps are stored is implementation specific.  In the
case of sparse arrays, this is not an issue because the nonzero
entries do not possess a natural order, unlike dense arrays in which
lexicographic ordering is used.  For multivariate polynomials, the
order of storage is not important algebraically because addition is
commutative and associative~\cite{hankin2022_disordR_arxiv}.

\subsection*{Compile-time options}

At compile time, the package offers two options.  Firstly one may use
the {\tt unordered\_map} class in place of the {\tt map}
class.  This option is provided in the interests of efficiency.  An
unordered map has lookup time~${\mathcal O}(1)$ (compare~${\mathcal
  O}(\log n)$ for the map class), but overhead is higher.

The other option offered is the nature of the key, which may be either
{\tt vector} class or {\tt deque} class.  Elements of a
{\tt vector} are guaranteed to be contiguous in memory, unlike a
{\tt deque}.  This does not appear to make a huge difference to
timings, but the default ({\tt unordered\_map} indexed by a
{\tt vector}) appears to be marginally the fastest option.

\end{document}